\documentclass[twocolumn,a4paper,fleqn]{article}
\usepackage{epsfig} 

\newtheorem{thm}{Theorem}[section]
\newtheorem{cor}[thm]{Corollary}

\newcommand{\be}{\begin{equation}}
\newcommand{\ee}{\end{equation}}
\newcommand{\ba}{\begin{eqnarray}}
\newcommand{\ea}{\end{eqnarray}}
\newcommand{\ban}{\begin{eqnarray*}}
\newcommand{\ean}{\end{eqnarray*}}

\newcommand{\ket}[1]{\mbox{$ | #1 \rangle $}}

\begin{document}

\title{\Large\bf Relativistic nonlocality (RNL)\\
in experiments with moving polarizers\\and 2 {\em non-before}
impacts}

{\normalsize{\author{{\bf Antoine Suarez}\thanks{suarez@leman.ch}\\
Center for Quantum Philosophy\\ The Institute for Interdisciplinary
Studies\\ P.O. Box 304, CH-8044 Zurich, Switzerland }}}
\maketitle

\begin{abstract}
RNL is a recently proposed relativistic nonlocal description, which
unifies relativity of simultaneity and superluminal nonlocality
(without superluminal signaling). In this article RNL is applied to
experiments with so-called 2 {\em non-before} impacts, leading to
new rules of calculating the joint probabilities, and predictions
conflicting with quantum mechanics. A real experiment using fast
moving polarizing beam-splitters is proposed.\\

{\em Keywords:}  relativistic nonlocality, timing-depen\-dent joint
probabilities, experiments with moving polarizers, 2 {\em
non-before} impacts.

\end{abstract}

\section{Introduction}

According to quantum mechanics (QM) there are two fundamental rules
of calculating the distribution of the outcomes in an experiment
\cite{feym65}: If one cannot distinguish (even in principle)
between different paths from source to detector, the amplitudes for
these alternative paths add coherently
(sum-of-probability-amplitudes rule, also called superposition
principle), and in multiparticle experiments nonlocal correlations
appear. If it is possible in principle to distinguish, the
probability of a determined outcome is the sum of the probabilities
for each alternative path (sum-of-probabilities rule), and in
multiparticle experiments the coincidence detections are correlated
according to local-realistic influences. This view implies that
indistinguishability is a sufficient condition for superposition,
and in particular for entanglement (multiparticle superposition)
\cite{ghz93}. The quantum formalism entails insensitivity to the
state of movement of the preparing and measuring devices.

In a previous article \cite{asvs97} it was argued that this view of
the relationship between entanglement and indistinguishability is
not the only possible one, and the principles of an alternative
nonlocal description were presented. According to it entanglement
depends not only on indistinguishability but also on the timing of
the impacts at the beam-splitters, and especially on the velocities
of these. Key notions are those of {\em before} and {\em
non-before} impacts at the beam-splitters. An experiments with 2
{\em before} impacts involving one fast moving polarizing
beam-splitter was proposed. Such an experiment may allow us to
decide between this alternative view and QM: Effectively, according
to QM the superposition rule applies because indistinguishability
happens, whereas according to the alternative description it is the
sum-of-probabilities rule that matters, because of the specific
timing involved.

In this article it is shown how the alternative description works
in a 2 {\em non-before} experiment: To account for the specific
timing it uses neither the sum-of-probabilities rule, nor the
sum-of-probability-amplitudes rule, but a new one. Thus the
proposed alternative description offers more possibilities to
calculate the distribution of the outcomes than does quantum
mechanics, and yields predictions conflicting with it.

Since relativity of simultaneity and (superluminal) nonlocality
(without superluminal signaling) are the basic features of the
proposed alternative description, it is called Relativistic
Nonlocality (RNL).

\section{Basic notions and principles of RNL}

We consider an experiment with entangled particle pairs in which
each beam-splitter can move fast, and change from one inertial
frame to another. The beam-splitters are labeled $BS_{i}$,
$i\in\{1,2\}$, and the corresponding detectors $D_{i}(+1)$ and
$D_{i}(-1)$.

The proposed RNL is based on the following definitions, principles
and rules:

\subsection{{\em Before} and {\em non-before} impacts}

If it is in principle impossible to know to which prepared
sub-ensemble a particle pair belongs by detecting each particle
after leaving the corresponding $BS_{i}$, then the impacts at the
splitters are referred to as originating indistinguishability or
uncertainty, and labeled $u_{i}$. If it is in principle possible to
know to which prepared sub-ensemble a particle pair belongs by
detecting each particle after leaving the corresponding $BS_{i}$,
then the impacts at the beam-splitters are referred to as making
possible distinguishability, and labeled $d_{i}$.

At the time $T_{i}$ at which a particle $i$ arrives at $BS_{i}$, we
consider whether in the inertial frame of this beam-splitter
particle $j$ ($j\in\{1,2\}, j\neq i$) has already made an impact at
$BS_{j}$ or not, i.e., whether $(T_{i}\geq T_{j})_{i}$ or
$(T_{i}<T_{j})_{i}$, the subscript $i$ after the parenthesis
meaning that all times referred to are measured in the rest frame
of $BS_{i}$. We introduce the following definitions:

\hspace*{5mm}

{\em Definition 1}: the impact of particle $i$ in $BS_{i}$ is a
{\em before} event $b_{i}$ if either $(T_{i}<T_{j})_{i}$, or the
impacts of the particles are $d_{1}$ and $d_{2}$ ones.

\hspace*{5mm}

{\em Definition 2}: the impact of particle $i$ in $BS_{i}$ is a
{\em non-before} event $a_{i}$ if:
\begin{enumerate}
\item{$(T_{i}\geq T_{j})_{i}$, and}
\item{the particles produce $u_{1}$ and $u_{2}$ impacts}
\end{enumerate}

\subsection{Measurable joint probabilities of coincidence counts
and unmeasurable conditional probabilities}

An experiment $e$ will be labeled by indicating the kind of impact
each particle undergoes, e.g. $e=(a_{1},b_{2})$.

A detection of particle $i$ is said to yield value $+1$ (or simply
value $+$) if the particle is detected in the detector $D_{i}(+1)$,
and value $-1$ (or simply value $-$) if the particle is detected in
$D_{i}(-1)$. A detection of a pair producing either outcome
$(+1,+1)$ or $(-1,-1)$ is said to yield total value $+1$ (or value
$+$); a detection of a pair producing either outcome  $(+1,-1)$  or
 $(-1,+1)$ is said to yield total value $-1$ (or value $-$).

Expressions like $Pe_{\sigma\omega}$, $\sigma,\omega\in\{+,-\}$,
denote the probabilities to obtain the indicated coincidence
detection values in experiment $e$ (i.e., particle 1 is detected in
$D_{1}(\sigma)$, particle 2 in $D_{2}(\omega)$). $Pe_{\sigma}$
denote the probability to obtain the total detection value $\sigma$
in experiment $e$. In a similar way, we write
$P^{QM}e_{\sigma\omega}$ for the probabilities predicted by
standard QM for experiment $e$ (note that in this case the impacts
will be referred to only by $u_{i}$ or $d_{i}$, since QM doesn't
consider differences in timing). The $Pe_{\sigma\omega}$ quantities
can be evaluated directly by measuring the corresponding count
rates in the corresponding experiment.

Further we denote by
$P\Big((a_{i})_{\sigma'}|(b_{i},b_{j})_{\sigma\omega}\Big)$ the
probability that a particle pair that would have produced the
outcome $({\sigma,\omega})$ in a $(b_{i},b_{j})$ experiment,
produces the outcome $({\sigma',\omega})$ if the experiment is a
$(a_{i},b_{j})$ one. Evidently, these conditional probabilities
cannot be evaluated from count rates, because if experiment
$(b_{i},b_{j})$ is performed on a determined particle pair, then it
is no longer possible to perform experiment $(a_{i},b_{j})$ on the
same pair. However, as we will see in section 3, RNL allow us to
establish rules calculating conditional probabilities from
measurable quantities.

\subsection{The nonlocal links behind the correlations}

Bell experiments with time-like separated impacts at the splitters
have already been done \cite{rar94}, demonstrating the same
correlations as for space-like separated ones. Consider an
experiment in which the choice particle $2$ makes in $BS_{2}$ lies
time-like separated after the choice particle $1$ makes in
$BS_{1}$. It is clear that at the time particle $1$ makes its
choice, it cannot account for choices in $BS_{2}$ because such
choices do not exist at all, from any observer's point of view. In
this case expressions like 'the later choice', and 'the former
choice' make the same sense in every inertial frame. Therefore, it
is reasonable to assume that the correlations appear because
particle 1 has to choose as it would choose in the absence of
nonlocal influences, and the choice particle 2 makes, depends
somewhat on the choice particle 1 has made.

Inspired by this explanation we introduce now the following
principles to account for the correlations when the impacts lay
space-like separated:

\hspace*{5mm}

{\em Principle I}: if the impact of a particle $i$ at $BS_{i}$ is a
$b_{i}$ impact, then particle $i$ produces values taking into
account only local information, i.e., it does not become influenced
by the parameters particle $j$ meets at the other arm of the
setup.\\

Accordingly:
\ba
P(b_{1},b_{2})_{\sigma\omega}=P^{QM}(d_{1},d_{2})_{\sigma\omega}.
\label{eq:2b}
\ea

\hspace*{5mm}

{\em Principle II}: if the impact of a particle $i$ at $BS_{i}$ is
a $a_{i}$ impact, then particle $i$ takes account of particle $j$
in such a way that the values particle $i$ actually produces, and
the values particle $j$ produces in a $b_{j}$ impact are correlated
according to the standard quantum mechanical entanglement rules.\\

Therefore, the following correlation rule must hold:

\ba
P(a_{1},b_{2})_{\sigma\omega}=P(b_{1},a_{2})_{\sigma\omega}\nonumber\\
=P^{QM}(u_{1},u_{2})_{\sigma\omega}.
\label{eq:1b1nb}
\ea

Notice that in all interference experiments performed till now both
splitters were at rest, and one of the impacts did happen always
before the other. Eq. \ref{eq:1b1nb} guarantees that RNL yields the
same predictions than QM for all experiments already done.

\hspace*{5mm}

{\em Principle III}: The choice particle $i$ makes does not take
into account the choice particle $i$ itself would have made if the
impact would have been a {\em before} one, i.e:

\ba
P\Big((a_{i})_{\sigma'}|(b_{i},b_{j})_{\sigma\omega}\Big)\nonumber\\
= P\Big((a_{i})_{\sigma'}|(b_{i},b_{j})_{(-\sigma)\omega}\Big)\nonumber\\
=P\Big((a_{i})_{\sigma'}|(b_{j})_{\omega}\Big).
\label{eq:cpr}
\ea

As argued in  \cite{asvs97}, by assuming {\em Principle I} and {\em
Principle II} we implicitly discard {\em firstly} the hypothesis
that the values produced by a particle, say particle 1, do depend
on the state of movement of the detectors $D_{1}$ monitoring
$BS_{1}$, and {\em secondly} the hypothesis that the values
produced by particle 1 depend on the time at which particle 2
impacts at a detector $D_{2}$, monitoring $BS_{2}$.

Suppose now that both impacts are {\em non-before} events. It would
be absurd to assume together that particle 1 chooses taking account
of the choice particle 2 has {\em really} made, and particle 2
chooses taking account of the choice particle 1 has {\em really}
made. Therefore we assume that particle 1 makes its choice in
$BS_{1}$ taking into account the choice particle 2 would have made
in $BS_{2}$ if the impact at this beam-splitter would have been a
{\em before} event, but that this choice of particle 1 is
independent of the choice particle 2 makes in the actual {\em
non-before} impact. Similarly particle 2 makes its choice in
$BS_{2}$ depending on which choice particle 1 would have made in
BS1 if the impact at this beam-splitter would have generated a {\em
before} event, but independently of the choice particle 1 makes in
the actual {\em non-before} impact.

These assumptions can be expressed through the following key
equation:

\hspace*{5mm}

{\em Principle IV:}

\ba
P(a_{1},a_{2})_{\sigma'\omega'}=\sum_{\sigma,\omega}
P(b_{1},b_{2})_{\sigma\omega}\nonumber\\
\times P\Big((a_{1})_{\sigma'}|(b_{2})_{\omega}\Big)P\Big((a_{2})_{\omega'}|(b_{1})_{\sigma}\Big).
\label{eq:2nb}
\ea

\hspace*{5mm}

According to Eq.(\ref{eq:2b}), in a 2 {\em before} experiment the
joint probabilities of coincidence detections are calculated
through sum-of-probabilities. According to Eq. (\ref{eq:1b1nb}) in
a 1 {\em before} 1  {\em non-before} experiment the joint
probabilities of coincidence detections are calculated through
sum-of-probabilities-amplitudes. In Section 3 it is shown that in
certain experiments with 2 {\em non-before} impacts Eq.
(\ref{eq:2nb}) yields a new rule that involves together
sum-of-probabilities and sum-of-probability-amplitudes.

It is worthy to highlight that RNL always involves instantaneous
influences even when the outcomes distribution is calculate through
sum-of-probabilities.

\subsection{Impossibility of communication without observable connection (signaling)}

RNL assumes further

\hspace*{5mm}

{\em Principle V:}

\be
P(a_{i})_{\sigma}=P(b_{i})_{\sigma}.
\label{eq:icws}
\ee

\hspace*{5mm}

The physical meaning of Eq. (\ref{eq:icws}) is the following: a
human agent at place A cannot produce observable order (a message)
at place B, if there is no observable connection (signaling)
between A and B. Accordingly, communication between (time-like or
space-like) separated human observers requires energy propagating
in space-time from one observer to the other. Indirectly this
principle leads also to the impossibility of using nonlocality for
superluminal signaling. Notice, however, that the principle works
also in situations with time-like separated measurements as for
instance when the impact at $BS_{2}$ lies time-like separated after
the detection at one of $D_{1}$. In such experiments interference
fringes at the level of the single detection (first-order
correlations) would not imply any superluminal signaling. However,
since there is no observable connection or signaling between any
$D_{1}$ and $BS_{2}$, first order interference fringes would imply
subluminal signal-less communication (i.e. the possibility of using
energyless or unobservable connections for generating observable
order). The motivation of Eq. (\ref{eq:icws}), therefore, is
primarily not the concern of limiting the speed of signaling, but
rather to forbid communication without signaling.

The impossibility of superluminal signaling is in physics a
consequence of the dependence of simultaneity on the inertial frame
(or the impossibility of absolute time) resulting from observations
like those of Michelson-Morley. This relativity of simultaneity
(and therefore the impossibility in principle of superluminal
signaling) enters evidently into RNL through the definition of {\em
before} and {\em non-before} impacts.

QM is a "specifically nonrelativistic" theory \cite{ghz93}, and the
concern of forbidding faster-than-light communication between human
observers was foreign to its construction. That quantum formalism
conspires to combine nonlocality with the impossibility of
superluminal signaling \cite{poro97} has the appearance of a "deep
mystery" \cite{ghz93} making possible a "pacific coexistence"
\cite{absh97} between quantum mechanics and relativity. The spirit
of RNL is somewhat the reverse: (superluminal) nonlocality, the
impossibility of communication without signaling, and constant c as
the upper limit for signaling are considered from the beginning to
be fundamental principles of the physical reality, and the
formalism has to adapt to them. They determine in particular the
path amplitudes and the possible correlations rules \cite{as97}.

\section{Theorems}

We derive now theorems that allow to calculate conditional
probabilities and the correlation coefficients in experiments with
2 {\em non-before} impacts.

\begin{thm}
\ba
P\Big((a_{i})_{\sigma}|(b_{j})_{\omega}\Big)=
\frac{P(a_{i},b_{j})_{\sigma\omega}}{P(b_{j})_{\omega}}
\label{eq:cpth}
\ea
\end{thm}

{\em Proof}: Conditional probabilities are related to measurable
quantities through the equation:

\ban
P(a_{i},b_{j})_{\sigma\omega} =
P(b_{i},b_{j})_{\sigma\omega}P\Big((a_{i})_{\sigma}|(b_{i},b_{j})_{\sigma\omega}\Big)\nonumber\\
+P(b_{i},b_{j})_{(-\sigma)\omega}P\Big((a_{i})_{\sigma}|(b_{i},b_{j})_{(-\sigma)\omega}\Big)
\ean

Taking into account (\ref{eq:cpr}) one is led to

\ban
P(a_{i},b_{j})_{\sigma\omega} =
\Big[P(b_{i},b_{j})_{\sigma\omega}+P(b_{i},b_{j})_{(-\sigma)\omega}\Big]\nonumber\\
\times P\Big((a_{i})_{\sigma}|(b_{j})_{\omega}\Big)\nonumber\\
= P(b_{j})_{\omega}P\Big((a_{i})_{\sigma}|(b_{j})_{\omega}\Big)
\ean

q.e.d.\\

For "maximally entangled states" in which the particles are
prepared equally distributed in two classes of pairs, it holds
that:

\ba
P(b_{i})_{\sigma}=P(b_{i})_{-\sigma}=\frac{1}{2}P(a_{i},b_{j})_{\sigma\omega}\nonumber\\
=P(a_{i},b_{j})_{{(-\sigma)}{(-\omega)}}
\ea

Then from Eq. (\ref{eq:cpth}) follows:\\

\begin{cor}
\ba
2P(a_{i},b_{j})_{\sigma\omega}=2P(a_{i},b_{j})_{{(-\sigma)}{(-\omega)}}\nonumber\\
=P\Big((a_{i})_{\sigma}|(b_{j})_{\omega}\Big)\nonumber\\
=P\Big((a_{i})_{-\sigma}|(b_{j})_{-\omega}\Big)
\label{eq:cpmet}
\ea
\end{cor}

\hspace*{5mm}

\begin{thm}
For experiments with particles prepared equally distributed in two
classes of pairs it holds that:

\be
E(a_{1},a_{2})=E(b_{1},b_{2})E(a_{1},b_{2})E(b_{1},a_{2}),
\label{eq:2nbth}
\ee

where each $E(e)$ denotes the correlation coefficient
$E=\sum_{\sigma,\omega=\pm 1}\sigma\omega P(e)_{\sigma\omega}$

\end{thm}

{\em Proof}: Expanding in $E(a_{1},a_{2})$ each of the four
$P(a_{1},a_{2})_{\sigma\omega}$ terms according to (\ref{eq:2nb})
yields:

\ban
E(a_{1},a_{2})=
\sum_{\sigma,\omega=\pm 1}\sigma\omega P(b_{1},b_{2})_{\sigma\omega}\nonumber\\
\times\Big[P\Big((a_{1})_{\sigma}\Big|(b_{2})_{\omega}\Big)-
P\Big((a_{1})_{-\sigma}\Big|(b_{2})_{\omega}\Big)\Big]\nonumber\\
\times\Big[P\Big((a_{2})_{\omega}\Big|(b_{1})_{\sigma}\Big)-
P\Big((a_{2})_{-\omega}\Big|(b_{1})_{\sigma}\Big)\Big]
\ean

Applying (\ref{eq:cpmet}) leads to:

\ban
E(a_{1},a_{2})=
\Big(\sum_{\sigma,\omega=\pm 1}\sigma\omega P(b_{1},b_{2})_{\sigma\omega}\Big)\nonumber\\
\times\Big[P\Big((a_{1})_{\sigma}\Big|(b_{2})_{\sigma}\Big)-
P\Big((a_{1})_{-\sigma}\Big|(b_{2})_{\sigma}\Big)\Big]\nonumber\\
\times\Big[P\Big((a_{2})_{\omega}\Big|(b_{1})_{\omega}\Big)-
P\Big((a_{2})_{-\omega}\Big|(b_{1})_{\omega}\Big)\Big]\nonumber\\
=\Big(\sum_{\sigma,\omega=\pm 1}\sigma\omega P(b_{1},b_{2})_{\sigma\omega}€\Big)\nonumber\\
\times\Big[2P(a_{1},b_{2})_{++} - 2P(a_{1},b_{2})_{-+}\Big]\nonumber\\
\times\Big[2P(a_{2},b_{1})_{--} - 2P(a_{2},b_{1})_{+-}\Big]\nonumber\\
=\Big(\sum_{\sigma,\omega=\pm 1}\sigma\omega P(b_{1},b_{2})_{\sigma\omega}\Big)\nonumber\\
\times\Big(\sum_{\sigma,\omega=\pm 1}\sigma\omega P(a_{1},b_{2})_{\sigma\omega}\Big)\nonumber\\
\times\Big(\sum_{\sigma,\omega=\pm 1}\sigma\omega P(b_{1},a_{2})_{\sigma\omega}\Big)
\ean
q.e.d.\\

Notice that Eq. (\ref{eq:2nbth}) is the product of the expression
$E(b_{1},b_{2})$ that works as the quantum mechanical
sum-of-probabilities, and the expressions $E(a_{1},b_{2})$ and
$E(b_{1},a_{2})$ that work as quantum mechanical
sum-of-probability-amplitudes.

\section{Predictions: Conflict (and agreement) between RNL and QM}

Consider now a 2 {\em non-before} experiment with entangled
polarized photons in which two classes of photon pairs,
$(H_{1},H_{2})$ and $(V_{1},V_{2})$, are prepared through
down-conversion in the "Bell state":

\be
\ket{\phi}=\frac{1}{\sqrt{2}}(\ket{H_{1},H_{2}}-\ket{V_{1},V_{2}})
\label{best}
\ee

where $H$ and $V$ indicate horizontal and vertical polarization,
respectively. The polarizing beam-splitters ($BS_{1}$ and $BS_{2}$)
are vertical oriented, and preceded by half wave plates, which
rotate the polarization of the photons by angles $\alpha$, $\beta$.

As said, the quantum formalism does not depend at all on the timing
of the impacts of the particles at the beam-splitters. The
correlation coefficient is assumed to be given by the
Lorentz-invariant expression \cite{bel64,agr82}:

\ba
E^{QM}=\sum_{\sigma,\omega}\sigma\omega
P^{QM}(u_{1},u_{2})_{\sigma\omega}\nonumber\\
=\cos 2(\alpha+\beta).
\ea

Consequently, for $\alpha+\beta=0$, QM predicts perfectly
correlated results (either both particles are transmitted, or they
are both reflected)for an experiment with 2 {\em non-before}
impacts.

Substituting (\ref{eq:2b}) and (\ref{eq:1b1nb}) in equation
(\ref{eq:2nbth}) leads to the correlation coefficient:

\be
E=\cos 2\alpha\cos 2\beta\cos^2 2(\alpha+\beta).
\label{eq:2nbph}
\ee

Consequently, according to RNL, $\alpha+\beta=0$ will not produce
$E=1$, i.e. perfectly correlated results. In particular, for
$\alpha=-\beta=45^{\circ}$  one gets $E=0$, i.e. the four possible
outcomes $(+1,+1)$, $(+1,-1)$, $(-1,+1)$, $(-1,-1)$ equally
distributed.

In summary, for $\alpha=-\beta=45^{\circ}$ QM and RNL lead to the
clearly conflicting predictions as well for the 2 {\em before}
experiment described in \cite{asvs97}, as for the 2 {\em
non-before} one.

Notice however that for $\alpha=\beta=22.5^{\circ}$, RNL and QM
lead to the same predictions in case of a 2 {\em non-before}
experiment, but to conflicting predictions in case of a 2 {\em
before} one.

\begin{figure}[t]
\centering\epsfig{figure=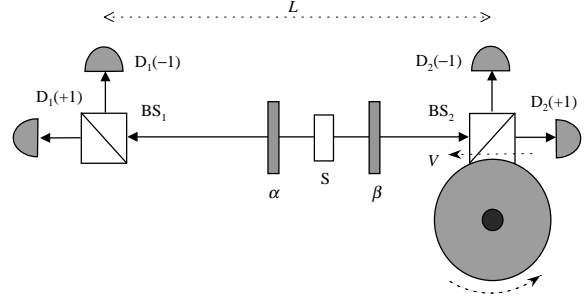,width=76mm}
{\small\it{\caption{Setup of the experiment with 2 {\em non-before}
impacts: Beam-splitter $BS_{1}$ at rest in the laboratory frame,
and beam-splitter $BS_{2}$ moving fast towards $BS_{1}$ at the
arrival of photon 2.}}}
\label{fig:setup}
\end{figure}

\section{En route towards a real relativistic nonlocality experiment}

A real experiment with 2 {\em non-before} impacts could be done
with the same setup of the 2 {\em before} experiment. The
arrangement is represented in Fig. 1. It is assumed that the
particles are channeled from the source to the beam-splitters by
means of optical fibers, and that the optical path $S$-$BS_{1}$
traveled by particle 1, is a bit longer than optical path
$S$-$BS_{2}$ traveled by particle 2. The delay in time resulting
from this path difference is labeled $\delta t$. Beam-splitter
$BS_{1}$ is at rest in the laboratory frame, and therefore the
impact at this splitter is a {\em non-before} one. Beam-splitter
$BS_{2}$ is set on a wheeler so that at the moment of the photon
impact it moves with velocity $V$ towards $BS_{1}$ (i.e. the
opposite direction as in the 2 {\em before} experiment). The
distance between the beam-splitters is labeled $L$.

The condition, to ensure a {\em non-before} impact at $BS_{2}$
follows straightforward from the analysis done in \cite{asvs97},
and is given by the equation:

\be
\delta t<\frac{VL}{c^2}.
\label{eq:cond}
\ee

Latest results confirm that values of $L$ greater than 100 km may
become possible in a couple of years \cite{tbg97}. This would mean
for velocities of about 100 m/sec (360 km/h) that $\delta t$ could
reach values of 111 ps. Such an accuracy in measurement does not
seem to be an insurmountable challenge, especially if the
increasing interest on quantum optics reaches the levels which
particle physics enjoys at present, and yields comparable budgets.

In conclusion: While RNL formulates its principles taking account
explicitly of the relativity of simultaneity, QM does not worry
about it. Thus while QM bears only two rules to calculate the joint
probabilities of coincidence detections, RNL can generate
additional ones and account for different timings. And so whereas
both pictures agree for all the experiments already done, they lead
to conflicting predictions regarding relativistic nonlocality
experiments. All this seems to suggest that the basic character of
QM may be that of an application of RNL to particular situations.
Anyway the fact that RNL unifies consistently relativity of
simultaneity and superluminal nonlocality speaks in favor of
continuing the effort to do the proposed relativistic experiments.
In particular it seems worthwhile to study whether satellites could
allow us to perform the experiment with much higher velocities
\cite{rar97}. More deeper analysis to clarify whether the
experiment is possible even with values of $L$ similar to those
used in \cite{tbg97} is in progress too \cite{gtz97}. The suggested
possibility of 2 {\em non-before} experiments through particles
impacting successively at 2 beam-splitters at rest \cite{as97}, is
further discussed in another article.

\section*{Acknowledgements}
I would like to thank Valerio Scarani (Ecole Polytechnique
F\'ed\'erale de Lausanne) for suggestions concerning the
formulation of RNL, Anton Zeilinger (University of Innsbruck) for
discussions about possible experimental realizations, and the Odier
and L\'eman Foundations for financial support.



\begin{thebibliography}{9}
\bibitem{rar97} This suggestion has been made by John Rarity and Paul Tapster in a private meeting.
\end{thebibliography}
\end{document}